\documentstyle[12pt,epsfig]{article}

\def\gcmm3{{\,{\rm g\,cm^{-3}}}}

\begin{document}


\begin{center}
\vglue .06in
{\Large \bf {Ultra High Energy Cosmic Rays:\\
propagation in the Galaxy and anisotropy
}}
\bigskip
\\{\bf O.E. Kalashev \footnote{e-mail: {\tt kalashev@ms2.inr.ac.ru}}},
 \\[.05in]
{\it{Institute for Nuclear Research of the Russian Academy of Sciences\\
60th October Anniversary Prospect 7a, Moscow 117312, Russia}}
\\[.05in]
{\bf Vadim A. Kuzmin \footnote{e-mail: {\tt kuzmin@ms2.inr.ac.ru}}},
\\[.05in]
{\it{Institute for Nuclear Research of the Russian Academy of
Sciences\\
60th October Anniversary Prospect 7a, Moscow 117312, Russia\\
and\\
Yukawa Institute for Theoretical Physics\\
Kyoto University, Kyoto 606-8502, Japan}}
\\[.05in]
{\bf D.V. Semikoz \footnote{e-mail: {\tt semikoz@mppmu.mpg.de}}}
 \\[.05in]
{\it{Max-Planck-Institut f\"ur Physik (Werner-Heisenberg-Institut)\\
F\"ohringer Ring 6, 80805 M\"unchen, Germany\\
and\\
Institute for Nuclear Research of the Russian Academy of Sciences\\
60th October Anniversary Prospect 7a, Moscow 117312, Russia}}
\\[.40in]
\end{center}


\begin{abstract}

We considered propagation of Ultra High Energy Cosmic Rays (UHECR) through
the galaxy.
We investigated models with sources of UHECR distributed
in the same way as Cold Dark Matter (CDM) in a self-consistent way, taking
into account both extra-galactic and Galactic contributions.
Using a very simple toy model of galactic magnetic field
we showed that in the case of galactic origin of UHECRs the anisotropy
can reach considerable values. 
In the case of extragalactic UHECRs origin
the anisotropy appears to be non-vanishing only for electron and photon
components due to synchrotron losses, but it hardly can be reassured.
The reason is an extremely low flux of UHE electrons and a too low level of
$\gamma$-ray anisotropy. \\
\\
\end{abstract}

\section{Introduction}

The observation of  Ultra High Energy Cosmic Ray (UHECR) events with
energies above $10^{20} eV$  ~\cite{fe1,agasa1} is one of the most intriguing
puzzles of modern physics. The point is that if UHECR come from large distances
their energy dramatically decreases due to interactions with background fields.
For example, nucleons of energies above $\simeq4\times10^{19} ~eV$ start to
interact with the Cosmic Microwave Background (CMB) and lose  their
energy due to pion
photoproduction (the so-called Greisen-Zatsepin-Kuzmin (GZK)
effect~\cite{GZK}). As a result, nucleons can not travel to an
observer from distances
larger than $R_{GZK} \simeq 50\, Mpc$. Heavy nuclei travel even much less
distances of about few Mpc~\cite{Puget} because they photodisintegrated by
CMB.
UHE electrons lose their energy due to synchrotron radiation in
extragalactic
magnetic fields. From other side, it is hard to  accelerate
protons, electrons and heavy nuclei up to such energies even in the most
powerful astrophysical objects~\cite{Hillas} such as radio
galaxies and active galactic nuclei, but even if it is possible,
usually such objects are situated far away from our galaxy on
distances larger than
$R_{GZK}$. This means that energy of a UHE particle will drop below the GZK
threshold unless its energy is larger than observable one on several
orders of magnitude.

It was suggested several ways to avoid this difficulty, but all of
them require either the introduction of a new physics well beyond
the Standard Model or extreme astrophysics.

One possible way to avoid GZK cutoff is to introduce a new particle, 
which interactions with CMB are suppressed in compare with the proton of the 
same energy.
The possible examples of such particles are light long-leaving 
gluino-containing baryons \cite{Farrar} or  axion-like particles \cite{axion}.

Another possibility is to stand with Standard Model particles but
change physics laws. For example, 
protons and neutrons still can be UHECR with energies larger then cutoff 
if one slightly violate Lorentz invariance \cite{VLI}.

One more way avoid a GZK cutoff is to produce UHE protons and photons 
inside of GZK volume from hidden sources. Photons with extremely high energies 
$E > 10^{22}$ eV have attenuation length larger then 100 Mpc, if Extragalactic
Magnetic Fields (EGMFs) are small $B_{\rm EGMF} < 10^{-11}$ G. Such photons
can be  the sources of the  secondary UHE photons inside of GZK sphere \cite{photons}.    
Another possible source of both UHE protons and photons are neutrinos
with $E > 10^{22}$ eV, which could interact with background neutrinos 
and produce secondary particles through $Z^0$ resonance \cite{weiler}.  
Both those scenarios require extreme astrophysical sources which should
accelerate primary photons up to energies $E > 10^{23}$ eV.

The other way to explain UHE events above the GZK cutoff is to create UHE
protons,
leptons and photons in decays of supermassive X-particles inside of
the GZK sphere. This is 
one of the so-called 'top-down' (TD) scenarios.
Depending upon the particular scenario, decaying X-particles could 
populate the Galaxy 
or can decay on cosmological distances \cite{berez1}. In the latter case the
important role is  played by
their decay secondaries' energy losses during propagation through the space.
The sources of the massive
X-particles themselves might be topological defects such as cosmic 
strings or magnetic
monopoles that could be produced in the early Universe during
symmetry-breaking phase transitions envisaged in Grand Unified
Theories \cite{berez2}. In an inflationary early Universe, the relevant topological
defects
could be formed at a phase transition at the end of
inflation.
Other possibility is to produce superheavy long-living particles thermally
during reheating epoch of the Universe~\cite{KuzRub} or
from vacuum fluctuations during or after inflation \cite{KuzTkach}.
For a review of TD models see, for example \cite{Sigl98}.

Important information about cosmic ray sources give the measurement of the UHECR 
arrival directions.
All experiments see uniform distribution of the UHECR over the sky.
This leads to the conclusion that most probably UHECR come from the
 extragalactic sources. Besides, AGASA experiment see small scale clusters 
\cite{clusters}. Those clusters can come from point-like sources \cite{corr}.  
Moreover, the correlations of UHECR with BL Lacertae was found \cite{BLLAC}.
If those small scale clusters and especially correlations with BL Lacs 
will confirmed in the future, TD models will be excluded.

The main goal of this work is a self-consistent consideration of the
galactic and extragalactic contributions to spectra of UHECR from TD models.
There are two major possibilities:  if  sources of UHECR are present in our
Galaxy or not.  In the first case simple estimates show that contribution
of the Galaxy component is about 100 times larger \cite{Peter98} than
extragalactic component and hence one might take into account the Galactic
contribution only. Our numerical calculations, based on detailed modeling
of UHECRs propagation process give even more strict limitations
on the ratio of the extragalactic component of UHECRs to the galactic one.

The main observational signature allowing to distinguish this kind of 
models is a quite significant anisotropy in
all the UHE particles spectra  which is in fact simply due to the
non-central position
of the Sun in the Galaxy at the distance $8.5~~ kpc$ from the Galaxy center.
In what follows  we present our results for
various models of UHECR source distributions in the Galaxy.

In the case there is no sources of UHECRs in the Galaxy, the anisotropy
in all particle spectra is vanishing, except for photons and electrons,
for which a small anisotropy in the directions to the Galactic center and
opposite one still exists due to synchrotron radiation.

\section{Propagation through the Galaxy}

A numerical codes has been developed recently for calculation of spectra
of UHECRs in the case of isotropic and homogeneous source and magnetic field
distributions \cite{Sigl98,Lee96,we1}. Our code provides a rather reliable
information on possible extragalactic sources of UHECRs. Unfortunately, as
was pointed earlier \cite{we1},
the method developed in above papers may not be used
directly if one is 
interested in sources within our Galaxy, moreover 
the galactic magnetic fields may not be neglected even for the
extragalactic component
of UHECRs in the case of light charged particles (electrons, positrons).
Electrons loose their energy rapidly enough due to synchrotron radiation in
the galactic magnetic field:

\begin{equation}
\frac{dE}{dt} = - \frac{4}{9}\frac{\alpha^2B^2E^2}{m_e^4},
\end{equation}
where $B \sim 10^{-6} G$ is the  galactic magnetic field, and $E$ is the
UHE electron energy.
Then the energy loss length is
\begin{equation}
 \left(\frac{1}{E}\frac{dE}{dt}\right)^{-1} =
3.8 \times
\left(\frac{E}{10^{15}eV}\right)^{-1}\left(\frac{B}{10^{-6}G}\right)^{-2}
kpc.
\end{equation}

Deflection is another important factor when dealing with the propagation
problem in a non-anisotropic case. The straight line propagation (SLP)
approximation which treats the motion of CR particles in one dimension
fails if the effect of the deflection becomes large. The gyroradius of a
charged particle with
charge $qe$ and momentum $p$ (energy $E$) is given by
\begin{equation}
R_g=\frac{p}{qeB_\perp} \simeq\frac{E}{qeB_\perp}
\simeq1.1\times\frac{1}{q}\left({E\over10^{18}{\,{\rm eV}}}\right)
\left({B_\perp\over10^{-6}{\rm G}}\right)^{-1}{\,{\rm kpc}}\,,
\end{equation}
where $B_\perp$ is the field component perpendicular to the
particle's momentum.

However, if one is  interested only in the spectra of particles with
energies larger than  $~10^{19}$~eV, i.e.  particles which weakly (up
to 10 degrees) deflect in the galactic magnetic fields
$B \simeq 10^{-6}$G on the distances $~10 kPc$,
one can use a straight-line propagation approximation.
This means that one can use the same kinematic equations
as in \cite{we1,Lee96} but written for the differential fluxes
instead of concentrations.
\begin{eqnarray}
\frac{d}{dt} j_a(E_a,t) = -j_a(E_a,t)\alpha_{a}(E_a,t) + \label{KinEq}\\
\sum_{c}\int\beta_{ca}(E_c,E_a,t)j_c(E_c,t)dE_c+\frac{Q_a(Ee,t)}{4\pi},
\nonumber
\end{eqnarray}
where  $j_{a}(E_a,t)$ is a differential flux of particles $a$.

Integration in Eq.~(\ref{KinEq}) is performed on the chosen straight
trajectory within the Galaxy. The initial values of $j_a(E)$ may
be taken from the previous task (extragalactic component). A some another 
difference is in a time dependence
of external source and magnetic field terms. Their values are
determined by current position within the trajectory.
Further, for simplicity we will assume spherical symmetry, that is
the source concentration
and the magnetic field strength are taken to be functions only of
distance from the center of the Galaxy $r$.
\begin{equation}
        B(t)=B^{\prime}(r(t)),~~~
        Q(t)=Q^{\prime}(r(t)),
\end{equation}
As a toy model of galactic magnetic fields we will use the one decreasing
exponentially with r:
\begin{equation}
B^{\prime}(r)=B_0~exp(-r/r_B). \label{Magn}
\end{equation}

\begin{figure}
\begin{center}
\rotatebox{0}{\epsfig{file=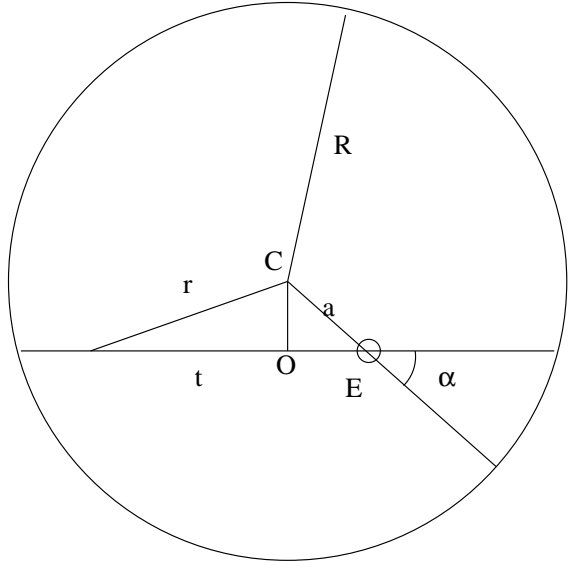,width=7.5cm,height=7.5cm}}
\end{center}
\caption{Propagation in the Galaxy. Points $C$ and $E$ show
locations of the galactic center and Earth respectively.}
\label{GalModel}
\end{figure}


 In the Fig.\ref{GalModel} we plot  point $E$, corresponding to the 
Earth location   at a distance $a$ from the galactic center $C$. 
Here $R$ is an effective Galaxy radius, the distance at which the source
concentration and
magnetic field strength become equal to extragalactic values.
Let us  draw an
arbitrary trajectory crossing point $E$ (horizontal line in the Fig.\ref{GalModel}), $\alpha$ being the angle
between the direction to the center and the trajectory and $t$ being
the length (time) parameter on the trajectory
(point $t=0$ is the nearest one to the center of Galaxy).

\begin{equation}
        r(t)=\sqrt{a^2sin^2\alpha+t^2},
\end{equation}
For the flux coming from "center" limits of integration are given by
\begin{equation}
        t_{min}=-\sqrt{R^2-a^2sin^2\alpha},
        t_{max}=\sqrt{a^2-a^2sin^2\alpha};
\end{equation}
for the flux from the opposite side
\begin{equation}
        t_{min}=\sqrt{a^2-a^2sin^2\alpha},
        t_{max}=\sqrt{R^2-a^2sin^2\alpha}.
\end{equation}

\section{Source distributions}

As we have already pointed out in Introduction, the absence or presence of
sources of UHECR in the Galaxy lead to quite different UHE
particle spectra on the Earth. In the first case only electron and photon
spectra are
slightly anisotropic, while in the latter case an anisotropy
is there for all particle spectra.

We considered the same
models of source distributions as
in \cite{Peter98}.
First 
is the isothermal halo model \cite{halo1}.
\begin{equation}
n(r)  \propto  {1\over (r_c^2 +r^2)}~,
\label{thermal}
\end{equation}
where $r_c$ is some core size. A model with the distribution (\ref{thermal})
may describe the collisionful CDM case.
The second model has the distribution of Ref.\cite{navar},
which corresponds to the opposite limiting case of collisionless CDM
\begin{equation}
n(r) \propto  {1\over \sqrt{(r_c^2 +r^2)} (R+r)^2}.
\label{nav}
\end{equation}

We will take always $r_c=2 kpc$ and $R=100 kpc$.

Since the  distribution of CDM in the Galaxy might contain most probably both
collisionful and collisionless components, the results for anisotropy
in a realistic case is probably  somewhere in between of two extreme cases
we considering here.

For the ratio of mean galactic and overall X-particle concentrations
we used value
\begin{equation}
\frac{<n_{X_{gal}}>}{<n_X>} =  \frac{1}{n_{gal}V_{gal}} = 10^5 .
\label{ratio}
\end{equation}

\section{Results}

The results of our numerical calculations of propagation of UHECR
through the Galaxy for the model of long-living X-particles \cite{KuzRub}
are presented on Figs.2-6. As an example, we used the following parameters
of the model: we took mass of a superheavy particle $m_X=10^{23} eV = 10^{14} GeV$.
We took as the main decay channel the $ X \rightarrow q q $ one.
We assumed that $X$-particles give the main contribution in the Cold Dark Matter 
(CDM) with $\Omega_X=0.3$ (assuming $\Omega_\Lambda=0.7$ and
$H_0=70 ~{\rm km} ~{\rm s}^{-1} {\rm Mpc}^{-1}$).
We normalized the resulting UHECRs flux to the observed one (AGASA data), 
and calculated required $t_X$ using the fact that fluxes $j_i \sim t_X^{-1}$.
Extragalactic magnetic fields were taken to be equal to $B_{out}=10^{-12} G$.
We used usual QCD-motivated fragmentation functions (see Eq.~ (52)
in ~\cite{Sigl98}).

We defined anisotropy in spectra of UHECR in the following simple way:
\begin{equation}
A = 2 \frac{J_{\rm center}-J_{\rm border}}{J_{\rm center}+J_{\rm border}}~, 
\label{anis}
\end{equation}
where $J_{\rm center}$ and $J_{\rm border}$ are the differential fluxes of UHECR from
the direction on the Galactic center and from opposite direction.

We considered two essentially different cases of the $X$-particle distribution. 
In  the first case we took uniform distribution of the $X$-particles
in the Universe, but without $X$-particles in our Galaxy. In the second case we 
considered   $X$-particle  distribution
similar to cold dark matter distribution in the galaxies. 

\begin{figure}
\begin{center}
\rotatebox{-90}{\epsfig{file=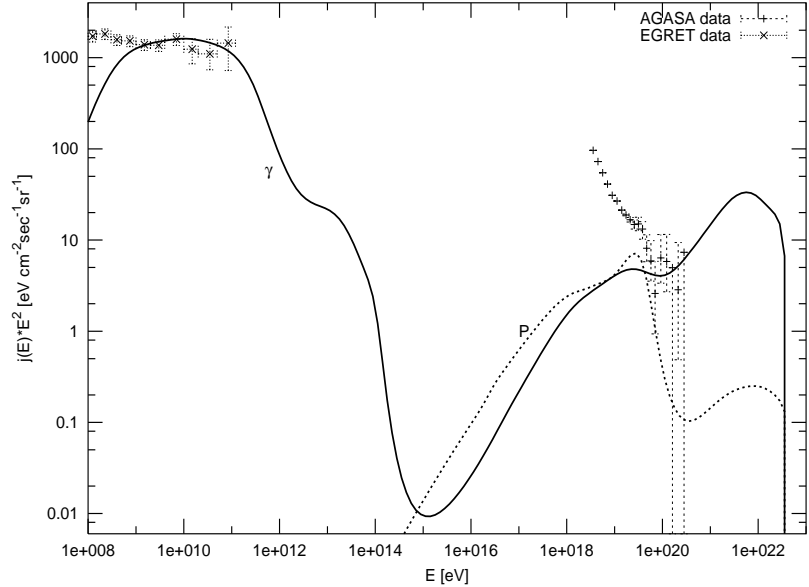,width=8cm,height=11cm}}
\end{center}
\caption{Spectra in the long-living X-particle model with
$\Omega_X=0.3$, $\Omega_\Lambda=0.7$, $H_0=70 ~{\rm km} ~{\rm s}^{-1} {\rm Mpc}^{-1}$,
$t_X = 8.6 \times 10^7 t_U$ ($t_U$
is the age of Universe) under constant EGMF of strength $B=10^{-12}$G,
assuming decay mode $X\rightarrow q q$.
This spectra is taken near the boundary of the Galaxy. 
The observed spectrum at energies more than $10^{19}$ eV will be similar, but without
electrons, which lose all the energy due to synchrotron radiation.
For this reason we don't show here the electron spectrum (in fact
it nearly coincides with photon one at high energies).
}
\label{no_gal_source}
\end{figure}
\begin{figure}
\begin{center}
\rotatebox{90}{\epsfig{file=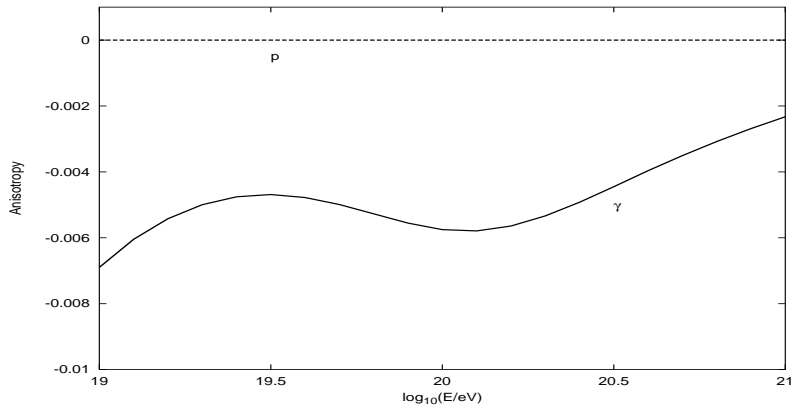,width=6.5cm,height=9cm}}
\end{center}
\caption{Anisotropy in spectra of photons due to synchrotron
radiation on galactic magnetic fields for the model of Fig.\ref{no_gal_source}.
Galactic magnetic field is taken in the form (\ref{Magn}), where
$B_0=10^{-6}$G, $r_B=4kpc$. There is no anisotropy in proton spectrum
in this case, because protons does not lose the energy.}
\label{anis_out}
\end{figure}
\begin{figure}
\begin{center}
\rotatebox{90}{\epsfig{file=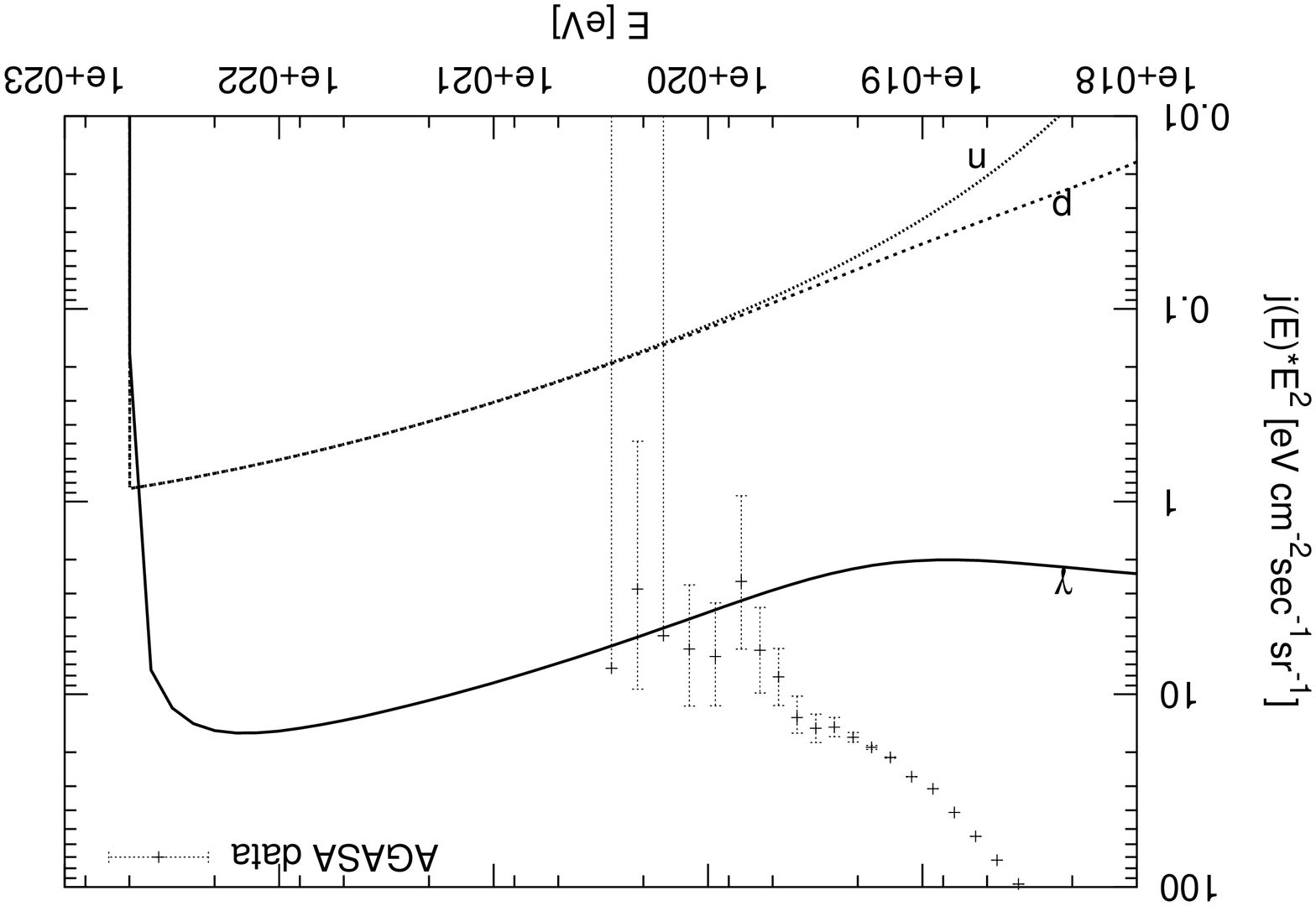,width=9.cm,height=10cm}}
\end{center}
\caption{Spectra of UHECR coming from Galaxy center in the case of (\ref{nav})
source distribution (similar to collisionless CDM distribution) in Galaxy.
$t_X = 5 \times 10^{11} t_U$. The electron flux is not shown here, it is
at least 100 times less then other fluxes at energies more then $10^{19}$eV. 
Flux of UHECR is dominated by photons, however another source of 
UHECR with energies below GZK cutoff is required.}
\label{navarro_sp}
\end{figure}

On the Fig.\ref{no_gal_source} we plot spectra of UHE protons and
photons near the boundary of our Galaxy in case of extragalactic
sources of UHECRs. Propagation through the Galaxy mostly leaves
the picture at highest energies without changes. It is not so only
for electrons, since they are strongly suppressed by galactic magnetic
field. Lifetime of $X$-particle in this case is 
$t_X = 8.6 \times 10^7 t_U$ ($t_U$ is the age of Universe).
From Fig.\ref{no_gal_source} one can conclude that all highest 
energy cosmic rays are photons. This is the general result for all TD models.
Moreover, for models with $m_X \geq few \times 10^{22}$ protons hardly can
contribute significantly in the region below cutoff, since an attempt
to increase extragalactic magnetic field to fit simultaneously highest
energy part of AGASA spectrum by photons
and lower energy part by protons will lead to overproducing of photons at
EGRET region, due to synchrotron radiation. And so in this case one need
to suppose that another kind of UHECR sources exists in order to explain
UHECR below cutoff.

The scale of spectra distortions produced during propagating through
the Galaxy can be illustrated in terms of the flux asymmetry in directions
to and from the Galaxy center Eq.(\ref{anis}). It is presented on the
Fig.\ref{anis_out}.
The spectra in any directions are similar for all particles, except
for photons: they are a little bit different due to influence
of galactic magnetic fields. Namely, some part of photons, which was converted
to electrons, was lost due to electrons synchrotron radiation in Galactic
magnetic field. Because the path through the Galaxy center is longer then from opposite direction, but 
both of them much smaller then photon  mean free path,
there is small asymmetry in photon flux, as we see from Fig.\ref{anis_out}.

At Fig. \ref{navarro_sp} we presented typical UHECRs spectra in the case of
the galactic origin. As was expected, the UHECR flux in this case is determined
by the source distribution rather than by propagation processes. This is not
the case only for electrons, because of the extremely small synchrotron energy
loss length in magnetic fields of the Galaxy.
The $X$-particle lifetime is $t_X = 1.5 \times 10^{11} t_U$.
We  checked that contribution of UHECR from all other galaxies 
is negligible in this case in compare to the contribution of the Milky Way.
The approximate relation for the Galactic versus extragalactic fluxes
is given by relation of corresponding $X$-particle lifetimes.
So, we get $F_{\rm Galaxy}/F_{\rm extragalactic} \sim 1.7 \times 10^3$.
As we see from  Fig. \ref{navarro_sp}, the spectrum of UHECR dominated
by photons at all energies. Protons contribute at most up $10\%$ to the flux.
This is simply due to fragmentation of quarks in $X$-particle decay.

\begin{figure}
\begin{center}
\rotatebox{90}{\epsfig{file=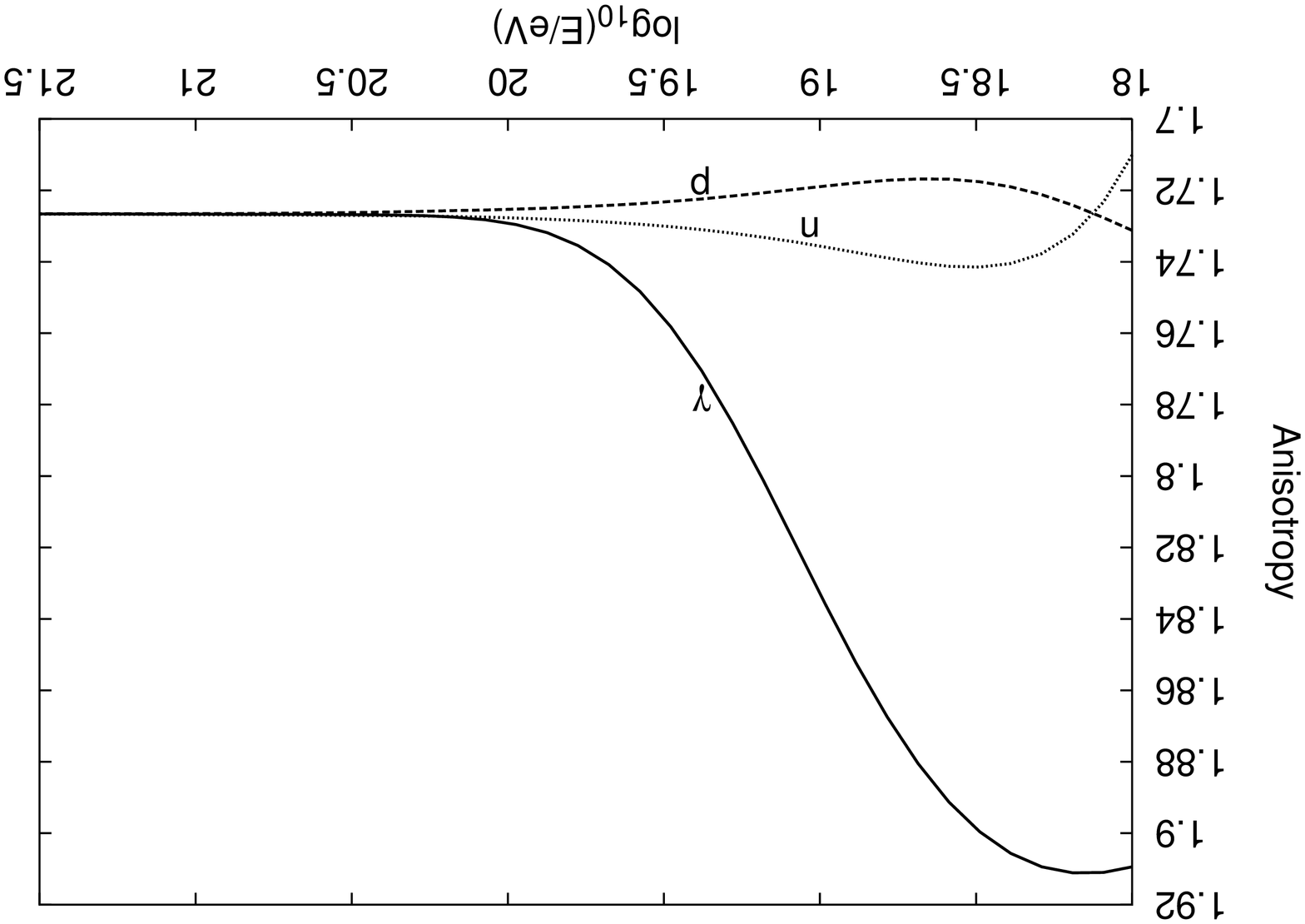,width=8.cm,height=10cm}}
\end{center}
\caption{Anisotropy in spectra of UHECR in directions to and from Galaxy center
for the thermal source distribution in Galaxy (collisionfull CDM).}
\label{thermal_anis}
\end{figure}
\begin{figure}
\begin{center}
\rotatebox{90}{\epsfig{file=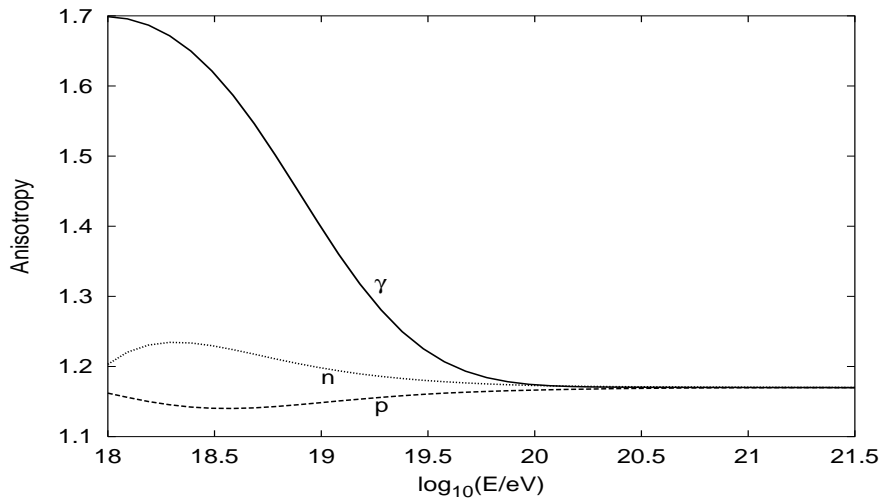,width=8cm,height=10cm}}
\end{center}
\caption{Anisotropy in spectra of UHECR in directions to and from Galaxy
center for the (\ref{nav}) source distribution (similar to collisionless CDM distribution) 
in Galaxy.}
\label{navar_anis}
\end{figure}

In the case of Galactic origin of UHECR the important parameter is anisotropy
in particle spectra, defined in  Eq.~(\ref{anis}). 
We investigated two essential cases 
of collisionless Eq.(\ref{nav}) and collisionful Eq.(\ref{thermal}) dark matter.
The asymmetry in particle spectra for those models
is plotted in Fig.\ref{thermal_anis} and  Fig.\ref{navar_anis}.

We conclude that anisotropy in spectra of nucleons, photons and neutrino
is an important test for top-down models, in which sources are located in
galaxies. In the realistic models of source distribution the anisotropy
can rise up to 80-90\%.
The dominant contribution in this case is provided by a flux coming
from our Galaxy. The extragalactic input in this case appears to be at
least 1000 times smaller.
The shape of the spectra at energies higher than few~$\times 10^{19}$
is determined mostly by source distribution. Deflection in galactic magnetic field
is negligible and so the exact configuration of the fields is not important.
The flux of UHECR will consist mostly from photons at highest energies.

In the case of the extragalactic UHECRs origin there will remain some 
anisotropy 
in photon and electron spectra, however extremely low 
flux of electrons and low value of photon anisotropy (less than 1\%)
make the anisotropy hardly detectable.
Nevertheless, it is probably worth to notice that in this case more
exact models of galactic magnetic fields may change the final result somewhat.

\section{Acknowledgments}
We thank S.L. Dubovsky,
 P.G. Tinyakov and I.I.Tkachev for fruitful discussion.
The work was supported in part by the Russian Fund for Fundamental Research grant
98-02-17493-a and INTAS grant 1A-1065.

\end{document}